\documentclass[aps,pra,twocolumn,showpacs,superscriptaddress]{revtex4-1}
\usepackage{graphicx}
\usepackage{amsmath,amsfonts}
\usepackage{bm}
\usepackage[usenames]{color}
\def\be{\begin{equation}}
\def\ee{\end{equation}}
\def\bea{\begin{eqnarray}}
\def\eea{\end{eqnarray}}
\def\bi{\begin{itemize}}
\def\ei{\end{itemize}}

\begin{document}

\title{Localizing spin dynamics in a spin-1 Bose-Einstein condensate via magnetic pulses}

\author{Huanbin Li}
\affiliation{School of Physics and Technology, Wuhan University, Wuhan, Hubei 430072, China}

\author{Zhengguo Pu}
\affiliation{School of Physics and Technology, Wuhan University, Wuhan, Hubei 430072, China}

\author{M. S. Chapman}
\affiliation{School of Physics, Georgia Institute of Technology, Atlanta, Georgia 30332-0430, USA}

\author{Wenxian Zhang}
\affiliation{School of Physics and Technology, Wuhan University, Wuhan, Hubei 430072, China}

\date{\today}

\begin{abstract}
Spin exchange interaction between atoms in a spin-1 Bose-Einstein condensate causes atomic spin evolving periodically under the single spatial mode approximation in the mean field theory. By applying fast magnetic pulses according to a two-step or a four-step control protocol, we find analytically that the spin dynamics is significantly suppressed for an arbitrary initial state. Numerical calculations under single mode approximation are carried out to confirm the validity and robustness of these protocols. This localization method can be readily utilized to improve the sensitivity of a magnetometer based on spin-1 Bose-Einstein condensates.
\end{abstract}

\pacs{67.85.-d,03.75.Kk,03.75.Mn }

\maketitle

\section{Introduction}

Spin exchange interaction between atoms in a spin-1 Bose-Einstein condensate (BEC) causes complex spin mixing dynamics and spin diffusion, which is a major obstacle to realize experimentally a high precision magnetometer based on spinor BEC~\cite{Chang05, Zhang05a, Sadler06, Vengalattore07, Abdullaev03, Liu09a, Steinke13, Stamper-Kurn13}. In order to improve the sensitivity of the magnetometer, a smaller spin exchange interaction is required, which may be implemented effectively by dynamical decoupling method using optical Feshbach resonance techniques~\cite{Zhang10a, Ning11}. In addition, the small spin exchange interaction can be utilized to resolve the ambiguity of the spin texture in ferromagnetically interacting $^{87}$Rb spin-1 BEC, where the spatial texture structure may be induced by the spin exchange interaction, the magnetic dipolar interaction, or both of them~\cite{Chang05, Sadler06, Vengalattore08, Vengalattore10, Kawaguchi10, Ning12, Eto14}.

However, a more experimentalist-friendly proposal to suppress the spin exchange interaction is employing the magnetic pulses and the microwave pulses, which are much easier to implement and tune experimentally~\cite{Gerbier06, Vengalattore08, Hoang13, Zhao14, Jiang14, Eto14a}. By applying a magnetic field to an atomic spin-1 BEC, only considered is the quadratic Zeeman effect $\delta$ which is proportional to the square of the field, because the linear Zeeman effect can be eliminated mathematically by adopting the rotating reference frame, due to the conservation of the total magnetization of the spin-1 condensate~\cite{Stenger98, Zhang03, Zhang05a}. Under current experimental conditions, the effective quadratic Zeeman energy of either the magnetic field or the microwave driving field can be adjusted from -240 Hz to +240 Hz, which is about 10 times larger than the spin exchange interaction for typical densities of a $^{87}$Rb spin-1 condensate, $\sim 10^{14}$ cm$^{-3}$~\cite{Zhang03, Hoang13, Zhao14}.

In this paper, we propose to localize the spin dynamics of a spin-1 BEC by periodically applying magnetic and/or microwave field pulses, which effectively suppress the spin exchange interaction. By applying two-step pulse cycles with positive $\delta$ only, the condensate dynamics is localized if the relative phase of the initial state is close to zero; by applying four-step pulse cycles with both positive and negative $\delta$, the condensate dynamics is localized for an {\em arbitrary} initial state. The exploration of the robustness of the protocols shows that a wide parameter regime exists for a spin-1 condensate under current experimental conditions. This proposal may find its potential application to improve the sensitivity of a practical high-resolution magnetometer based on spin-1 BEC.

The paper is organized as follows. In Sec.~\ref{sec:td}, we review the theoretical description of the free spin mixing dynamics under the single spatial mode approximation (SMA) in a spin-1 BEC in a magnetic field, whose quadratic Zeeman splitting $\delta$ ranges from large negative values to large positive values. In Sec.~\ref{sec:ld}, we analytically design and numerically confirm the control protocols of magnetic/microwave pulses to localize the condensate spin dynamics, where either a two-step or a four-step pulse cycle is employed. Furthermore, the robustness of the control protocols is explored in Sec.~\ref{sec:r} by assuming 5\% random error of the pulse amplitude $\delta(t)$. Finally, a brief summary is presented in Sec.~\ref{sec:con}.

\section{Free spin dynamics in a magnetic field}
\label{sec:td}

Within the mean field theory, the free spin mixing dynamics in a spin-1 BEC with either ferromagnetic or antiferromagnetic spin exchange interaction under the SMA in a magnetic field is described by the following equation of motion~\cite{Zhang05a, Zhang10a, Zhao14}
\begin{eqnarray}
\label{Ea}
\dot{\rho_0} &=& \frac{2c}{\hbar}\rho_0\sqrt{(1-\rho_0)^{2}-m^{2}}\sin \theta, \nonumber \\
\dot{\theta} &=& -\frac{2\delta}{\hbar}+\frac{2c}{\hbar}(1-2\rho_0) \\
&& +\frac{2c}{\hbar}\;\frac{(1-\rho_0)(1-2\rho_0)-m^{2}}{\sqrt{(1-\rho_0)^{2}-m^2}}\cos \theta, \nonumber
\end{eqnarray}
where $c=c_2N\int d \vec{r} |\phi(\vec{r})|^4$ with $N$ being the total number of atoms in the condensate and $\phi(\vec{r})$ a normalized spatial mode function under the SMA, which is determined by a scalar Gross-Pitaevskii equation with a spin-independent interaction, $[-(\hbar^2/2M)\nabla^2+V_{\rm ext}(\vec{r})+c_0|\phi|^2]\phi(\vec{r})= \mu \phi(\vec{r})$ where $M$ is the atomic mass and $V$ is the external harmonic trapping potential. The spin independent coefficient $c_0$ and spin exchange coefficient $c_2$ are given, respectively, by $c_0=4\pi\hbar^2(a_0+2a_2)/3M$ and $c_2=4\pi\hbar^2(a_2-a_0)/3M$ with the $s$-wave scattering length $a_0(a_2)$ for two spin-1 atoms in the compound symmetric channel of total spin 0(2). For two popular ultracold spin-1 atomic gases in experiments, $^{87}$Rb and $^{23}$Na, $c_0 \gg |c_2|$ is always satisfied and thus guarantees the validity of the SMA in most experimental situations~\cite{Law98, Yi02, Zhang03}.

The fractional population of spin component $\rho_\alpha (\alpha=-1,0,+1)$ satisfies $\sum_{\alpha} \rho_\alpha=1$. The magnetization $m=\rho_+-\rho_-$ is a constant during the evolution, due to the isotropic nature of the spin exchange interaction. The relative phase among the three components is $\theta = \theta_+ + \theta_- - 2\theta_0$ with $\theta_\alpha$ being the phase of the spin wave function. The quadratic Zeeman energy is  $\delta=(E_++E_--2E_0)/2$ with $E_\alpha$ the Zeeman energy shift of the component. In general, $\delta \approx 72 B^2$ Hz/G$^2$ for $^{87}$Rb BECs and $\delta \approx 278 B^2$ Hz/G$^2$ for $^{23}$Na BECs, where the magnetic field $B$ is in unit of Gauss. Due to the conservation of the magnetization $m$, the linear Zeeman energy $(E_--E_+)/2$ can be eliminated mathematically by adopting a rotating reference frame.

The total spin energy is a constant during the free evolution of the spin-1 condensate in a magnetic field
\begin{equation}  \label{EE}
\varepsilon=c\rho_0\left[(1-\rho_0)+\sqrt{(1-\rho_0)^{2}-m^{2}}\cos(\theta)\right]+\delta(1-\rho_0).\nonumber
\end{equation}
Starting from a given initial state, which is usually a ground state in a magnetic field in experiments, the condensate evolves according to an iso-energy trajectory in the plane of $\rho_0$-$\theta$, by changing abruptly the magnetic field to a different value. By taking into account of the energy conservation, the Eq.~(\ref{Ea}) is further simplified as $(\dot{\rho_0})^2=(4/\hbar^2)\{[\varepsilon-\delta(1-\rho_0)] [(2c\rho_0+\delta)(1-\rho_0)-\varepsilon]-(c\rho_0m)^2\}$
thus the time evolution of $\rho_0$ can be analytically expressed in terms of the Jacobian elliptic function cn(.,.) if $\delta \neq 0$ and sinusoidal function if $\delta = 0$~\cite{Pu99, Zhang05a, Zhang10a}，
\begin{eqnarray} \label{eq:1}
\rho_0(t)=\frac{1}{2}[x_2+x_1-(x_2-x_1)\sin(\gamma_0+2t\sqrt{2c\varepsilon+c^2m^2})]\nonumber\\
\end{eqnarray}
for $\delta = 0$;
\begin{eqnarray} \label{eq:1A}
\rho_0(t)&=& x_2+(x_3-x_2)cn^{2}(\gamma_0+t\sqrt{2c\delta(x_3-x_1)},k) \nonumber \\
\end{eqnarray}
for $c\delta >0$;
\begin{eqnarray}  \label{eq:1B}
\rho_0(t)&=& x_2-(x_2-x_1)cn^{2}(\gamma_0+t\sqrt{-2c\delta(x_3-x_1)},k) \nonumber \\
\end{eqnarray}
for $c\delta <0$. We have set $\hbar = 1$. Here $x_1\leq x_2\leq x_3$ ($x_1\leq x_2$ for $\delta=0$) are the roots of $\dot{\rho_0}=0$ , $k=\sqrt{(x_3-x_2)/(x_3-x_1)}$ if $c\delta >0$, and $k=\sqrt{(x_2-x_1)/(x_3-x_1)}$ if $c\delta <0$. $\gamma_0$ is determined by the initial state. Hereafter we
assume $c=-1$ thus the energy unit is $|c|$ and  the time unit is $|c|^{-1}$ and $m=0$.

\begin{figure}
\includegraphics[width=3.25in]{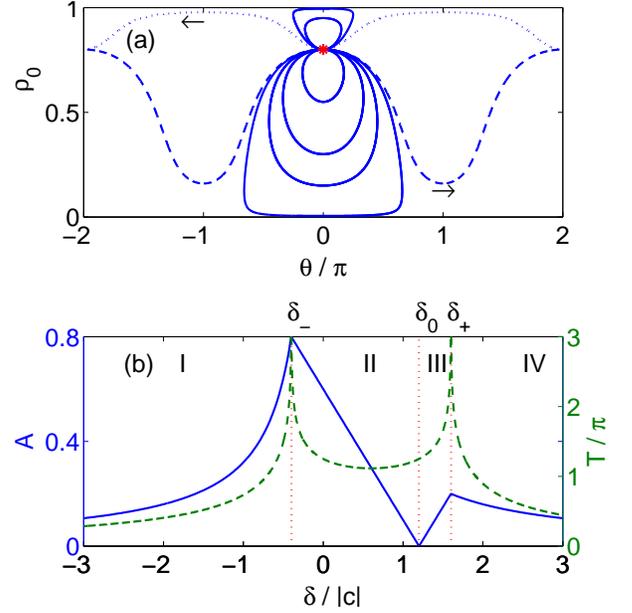}
\caption{(Color online) (a) Typical trajectories in $\rho_0$-$\theta$ plane for $\delta / |c|=-0.5$ (dashed line), $-0.39, -0.1, 0.2, 0.7, 1.5, 1.59$ (solid lines), $1.8$ (dotted line), from bottom to top. All the trajectories with solid lines evolve in a clockwise direction. The initial state (asterisk) is $\rho_{0i}=0.8$ and $\theta_i=0$. (b) Dependence of the oscillation amplitude $A$ (blue solid line) and the period $T$ (green dashed line) of $\rho_0$ on $\delta$. The running phase modes corresponds to the region I and IV and the oscillatory modes corresponds to region II and III. The red dotted lines marked by $\delta_0$ and $\delta_{\pm}$ denotes, respectively, the ground state quadratic Zeeman energy (the initial state coincides with the ground state and $A$ is zero) and the resonant quadratic Zeeman energy ($T$ is infinite and $\rho_0 \rightarrow 0$ or $1$).
}
\label{x_delta}
\end{figure}

Typical trajectories are illustrated in Fig.~\ref{x_delta}(a) for different $\delta$. Although starting from the same initial state, the trajectories could cover the whole $\rho_0$-$\theta$ plane by continuously varying the quadratic Zeeman energy $\delta$ from negative infinity to positive infinity. All the trajectories are classified into two modes: the oscillatory mode where $\theta$ is between $[-\pi,\pi]$ and the running phase mode where $\theta$ goes beyond $[-\pi,\pi]$. As shown obviously in Fig.~\ref{x_delta}(a), the oscillatory mode trajectories are evolving in a clockwise (counterclockwise) direction if $c<0$ ($c>0$), while the running phase mode trajectories for large $|\delta|\gg |c|$ may take one of two opposite directions, depending on the sign of $\delta$. This is a key point in order to localize the condensate spin dynamics. The boundaries between the oscillatory modes and the running phase modes satisfy one of the two requirements, $\rho_0(t) = 0$ ($\delta = \delta_-$) or $1$ ($\delta = \delta_+$) if time is long. The corresponding period $T$ becomes infinite [see also Fig.~\ref{x_delta}(b)]. Another special point $\delta = \delta_0$ denotes the coincidence of the initial state with the ground state thus the oscillation amplitude $A$ is zero but the period $T$ is finite.

The oscillation amplitudes and the periods are shown in Fig.~\ref{x_delta}(b). There are clearly four regions: (I) running phase mode with increasing $\theta(t)$; (II) oscillatory mode with $0<\rho_0(t) \le \rho_{0i}$; (III) oscillatory mode with $ \rho_{0i} < \rho_0(t) < 1$; (IV) running phase mode with decreasing $\theta(t)$. The amplitude of the oscillations $A$ monotonically increases in regions (I) and (III) but decreases in regions (II) and (IV) with $\delta$ increasing. The period of the oscillations $T$ shows two resonant peaks at $\delta = \delta_{\pm}$ where $\rho_0(t)=0$ or $1$ at long enough time. The period is almost a constant between these two peaks but decreases rapidly outside the peaks. Similar oscillation behaviors were observed also in antiferromagnetically interacting $^{23}$Na spin-1 condensates ($c>0$)~\cite{Zhao14, Jiang14}.

We observe from Fig.~\ref{x_delta}(a) that in the oscillatory mode $\theta$ increase or decrease with time if $\theta \approx 0$ and $\rho_0$ is around its extremes. We may utilize this property to localize the condensate dynamics around $\theta\approx 0$ by canceling $\theta$ in a period with $\theta$ increasing (decreasing) during the first (second) part. For an arbitrary state, however, we may utilize both the oscillatory and the running phase modes to localize the dynamics since $\theta$ may increase or decrease for a given $\rho_0$, depending on the value of $\delta$.

\section{Localized spin dynamics}
\label{sec:ld}

\begin{figure}
\includegraphics[width=3.25in]{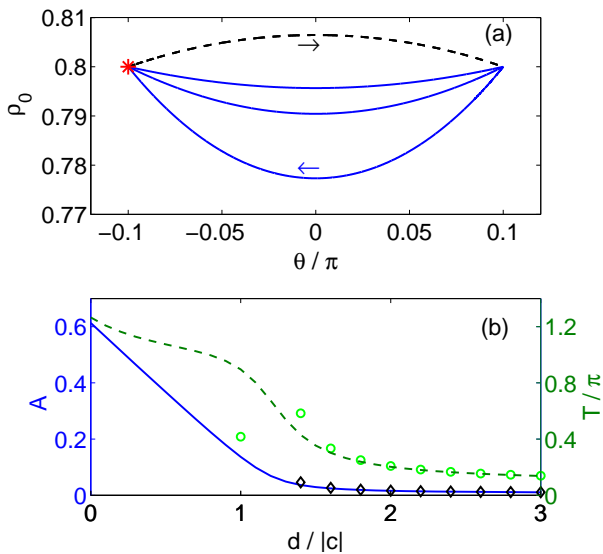}
\caption{(Color online) (a) Typical trajectories under the modulation of $\delta(t)$. Each modulation cycle includes a free evolution with $\delta = 0$ (dashed line) and a controlled evolution with $d / |c| = 1.5, 2.0, 3.0$ (solid lines, from bottom to top). The initial state (red asterisk) is $\rho_{0i}=0.8$ and $\theta_i=-0.1\pi$. (b) Dependence of the amplitude (blue solid line) and period (green dashed line) of $\rho_0$ on $d$, the nonzero quadratic Zeeman splitting. Circles and diamonds are the period and amplitude calculated analytically with Eqs.~(\ref{ApproT}) and~(\ref{ApproA}) for large $d$s, respectively.
}
\label{2step}
\end{figure}

We consider first that the control period consists of two steps, a free evolution ($\delta = 0$) for a time slot $\tau_1$ followed by an evolution in a magnetic field $\delta = d$ for a time slot $\tau_2$. We refer hereafter this protocol as two-step control. For a given initial state with $\theta_i \approx 0$, it is easy to prove analytically that $\tau_1$ depends on the initial state and $\tau_2$ depends uniquely on $d$, which indicates that there is only one free parameter in the two-step control protocol. The time dependence of the magnetic field for the two-step control is
\begin{equation}\label{deltat}
  \delta(t)=\left\{ \begin{array}{ll}
           0, \; & j(\tau_1+\tau_2)\leq t< j(\tau_1+\tau_2)+\tau_1 \\
           d, \; & j(\tau_1+\tau_2)+\tau_1\leq t< (j+1)(\tau_1+\tau_2)
           \end{array}\right. \nonumber
\end{equation}
where $j=0,1,2,\cdots$ is an integer denoting the number of control cycles.

Typical controlled trajectories are illustrated in Fig.~\ref{2step}(a) for three values of $d$, where the initial state is $\rho_{0i} = 0.8$ and $\theta_i = -0.1 \pi$.
We see clearly that the oscillations of both $\rho_0$ and $\theta$ under the two-step control are smaller than that during free evolution, indicating that the condensate dynamics is indeed localized by the two-step control protocol. Starting from the same initial state,  the lager the $d$ is, the smaller the oscillation of $\rho_0$ is.

The condensate spin average is $\langle \bm{F} \rangle = \langle F_x \rangle \hat{x}+\langle F_y\rangle\hat{y}+ \langle F_z\rangle \hat{z}$ for a state with~\cite{Zhang05a}
\begin{eqnarray}
\langle F_x\rangle + i\langle F_y\rangle &=& 2 \sqrt{\rho_0 (1-\rho_0)} \cos(\theta/2), \nonumber\\
\langle F_z\rangle &=& 0. \nonumber
\end{eqnarray}
Once we localize $\rho_0(t)$ and $\theta(t)$, the condensate spin $\langle \bm{F} \rangle$ is obviously localized. For a nonzero $\langle F_z \rangle$, the localization occurs similarly.

The cycle period $T$ depends on the free evolution time $\tau_1$ and the controlled evolution time $\tau_2$. The free evolution time is determined by the evolution time of the system from its initial state $\rho_{0i}$ and $\theta_i$ to the symmetric state $\rho_0(\tau_1) = \rho_{0i}$ and $\theta(\tau_1) = -\theta_i$. In this way, the time $\tau_1$ is calculated analytically by using Eq.~(\ref{eq:1})
\begin{eqnarray}
 \tau_1=(\pi/2-\gamma_0)/\sqrt{2\,c\,\varepsilon+c^2m^2}
\end{eqnarray}
In the limit of small $\theta \ll 1$, $\tau_1 \approx |\theta_i/[2c(1-2\rho_{0i})]|$ where we have used $\rho_0(t) \approx \rho_{0i}$. Similarly, the controlled evolution time $\tau_2$ is the evolution time of the system in the magnetic field $\delta = d$ and can be calculated, by using the conditions $\rho_0(T) = \rho_{0i}$ and $\theta(T) = \theta_i$,
\begin{eqnarray}
 \tau_2=\sqrt 2\gamma_0'/\sqrt{-c\,d (x_3-x_1)}
\end{eqnarray}
In the limit of small $\theta \ll 1$ and large $d$, $\tau_2 \approx |\theta_i/[2c(1-2\rho_{0i})-d]|$. In total, the cycle period is approximated as
\begin{equation}
\label{ApproT}
T \approx \left|\frac{\theta_i\,d}{2c(1-2\rho_{0i})[2c(1-2\rho_{0i})-d]}\right|
\end{equation}
for small $\theta_i$ and large $d$. We notice that $T\approx \tau_1$ for large $d$, as shown also in Fig.~\ref{2step}(b).

We define the control oscillation amplitude as $A = \max(\rho_0) - \min(\rho_0)$, which depends obviously on the initial state and the magnetic field $d$. The amplitude can be calculated analytically but is too lengthy to present here. In the limit of large $d$ and small $\theta_i$, the amplitude is approximately
\begin{equation}\label{ApproA}
A = A_1+A_2
\end{equation}
where
\begin{eqnarray*}
    A_1 & \approx & \frac{\rho_{0i}(1-\rho_{0i})}{4(2\rho_{0i}-1)} \; \theta_i^2, \\
    A_2 & \approx &  \frac{\rho_{0i}(1-\rho_{0i})} {|2d/c|-4(2\rho_{0i}-1)} \; \theta_i^2.
\end{eqnarray*}
We see that $A$ approaches to $A_1$ as $d$ goes to infinity.

In Fig~\ref{2step}(b), we present the dependence of $A$ and $T$ on the control magnetic field $d$. We see clearly that $A$ and $T$ decrease monotonically as $d$ increases, manifesting the fact that better localization of the condensate dynamics is achieved in a higher magnetic field. We note that $A$ and $T$ approach their {\em nonzero} asymptotic values at large values of $d$. Actually, to reduce the oscillation amplitude $A$ further down to zero, we have to employ the following four-step control protocol.

\begin{figure}
\includegraphics[width=3.25in]{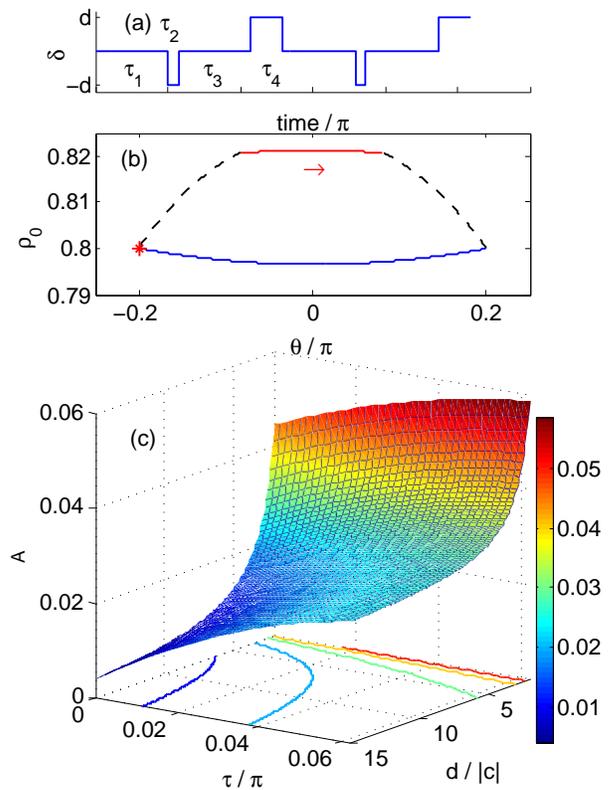}
\caption {(Color online) (a) Schematic of a magnetic pulse sequence. (b) Typical controlled trajectory of $\rho_0(t)$ and $\theta(t)$ under a four-step pulse sequence of $\delta(t)$ for a four-step protocol. The red asterisk marks the initial state. The parameters are $\rho_{0i}=0.8$, $\theta_i=-0.2\pi$, $d/|c|=10$, and $\tau_1 = \tau_3 = \tau=0.05\pi$. The values of $\tau_2$ and $\tau_4$ are calculated, $\tau_2\approx 0.0072\pi$ and $\tau_4 \approx 0.0226\pi$. (c) Amplitude of $\rho_0$ under four-step pulse sequences. Better localization of $\rho_0$ (smaller $A$) is achieved for larger $d$ and smaller $\tau$.}
\label{4step}
\end{figure}

We consider next that the control period consists of four steps, (i) a free evolution for a time $\tau_1$, (ii) an evolution in a magnetic field with $\delta = d_1$ for a time $\tau_2$, (iii) a second free evolution for the time $\tau_3$, and (iv) a second controlled evolution in another magnetic field with $\delta = d_2$ for a time $\tau_4$, as shown in Fig.~\ref{4step}(a). We refer this protocol as four-step control. For simplicity but without loss of generality, we limit to the symmetric situations where $d_1 = -d_2 = d$ and $\tau_3 = \tau_1 = \tau$. It will be analytically proved that $\tau_2$ and $\tau_4$ are uniquely determined by $d$ and $\tau$. Thus there are only two free parameters, $d$ and $\tau$, in the four-step control we considered here.

It is straightforward to find the analytical solution to $\tau_2$ and $\tau_4$, by using the initial state and the Eqs.~(\ref{eq:1}-\ref{eq:1B}),
\begin{eqnarray}
  \label{eq:4tau}
  \tau_2 &=& \frac{\sqrt 2 \gamma_0}{\sqrt{-cd(x_3-x_1)}},  \nonumber \\
  \tau_4 &=& \frac{\sqrt 2 \gamma_0'}{\sqrt{cd(x_3'-x_1')}} \nonumber
\end{eqnarray}
where $K(k)$ is the elliptic integral of the first kind. $\gamma_0'$ and $x_{3,1}'$ are determined by $\rho_0(\tau)$ and $-\theta(\tau)$. We note here that the initial state for $\tau_2$ is $\rho_0(\tau)$ and $\theta(\tau)$, and that for $\tau_4$ is $\rho_{0i}$ and $-\theta_i$. The total period for a complete cycle is
\begin{equation}\label{eq:4T}
    T = 2 \tau + \frac{\sqrt 2 \gamma_0}{\sqrt{-cd(x_3-x_1)}} + \frac{\sqrt 2 \gamma_0'}{\sqrt{cd(x_3'-x_1')}}.
\end{equation}

Typical controlled evolution of the condensate is illustrated in Fig.~\ref{4step}(b), where the parameters are given in the caption. Compared to the two-step control protocol, there are two advantages. The first is that the initial state is arbitrary, particularly, $\theta_i$ goes beyond the smallness requirement. The second is that the oscillation amplitude and period approach to zero if $d$ is large enough and $\tau$ is short enough, as shown in Fig.~\ref{4step}(c) and Eq.~(\ref{eq:4T}).

\section{Robustness of the control protocols}
\label{sec:r}

\begin{figure}
\includegraphics[width=3.25in]{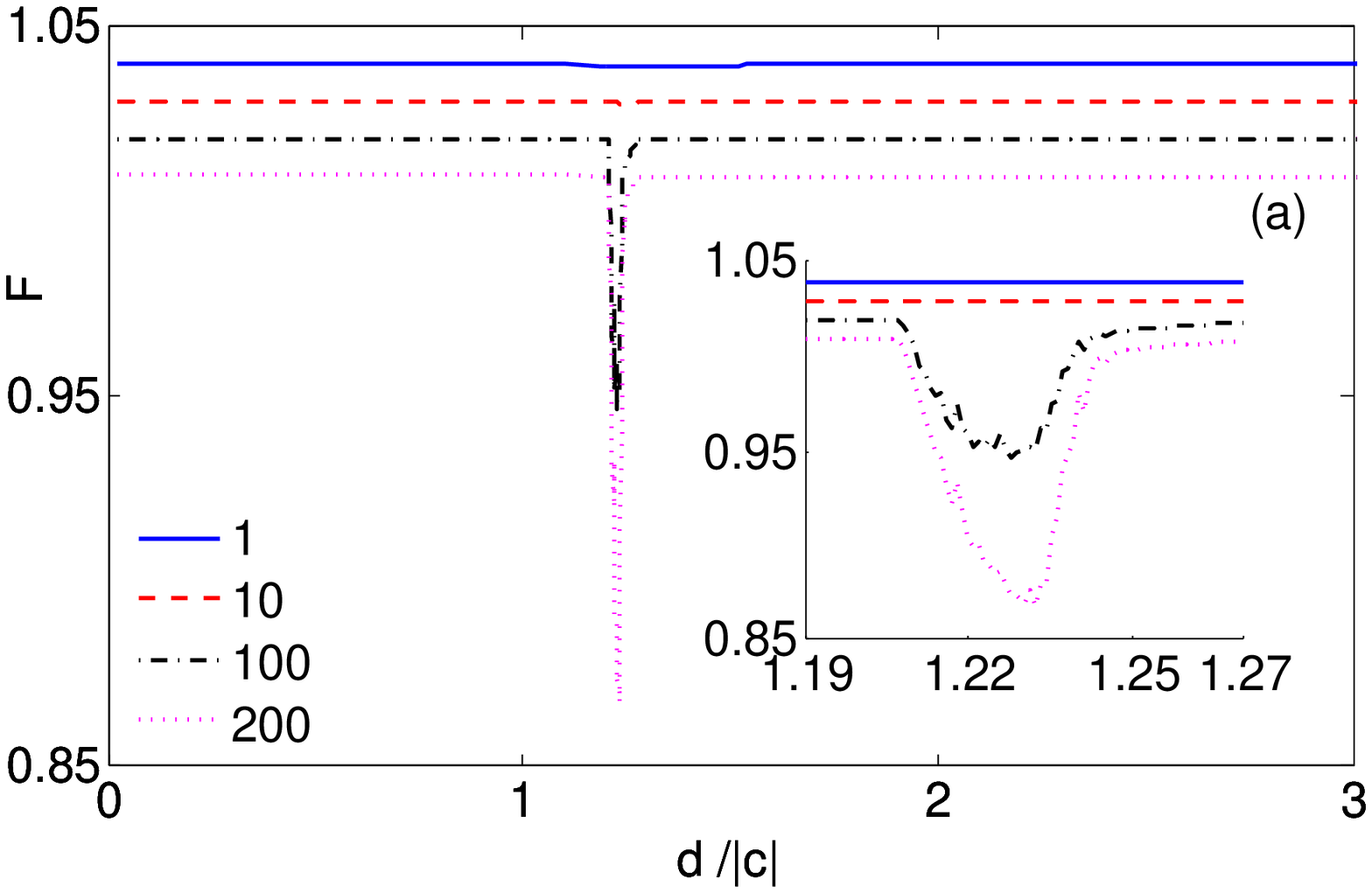}
\includegraphics[width=3.25in]{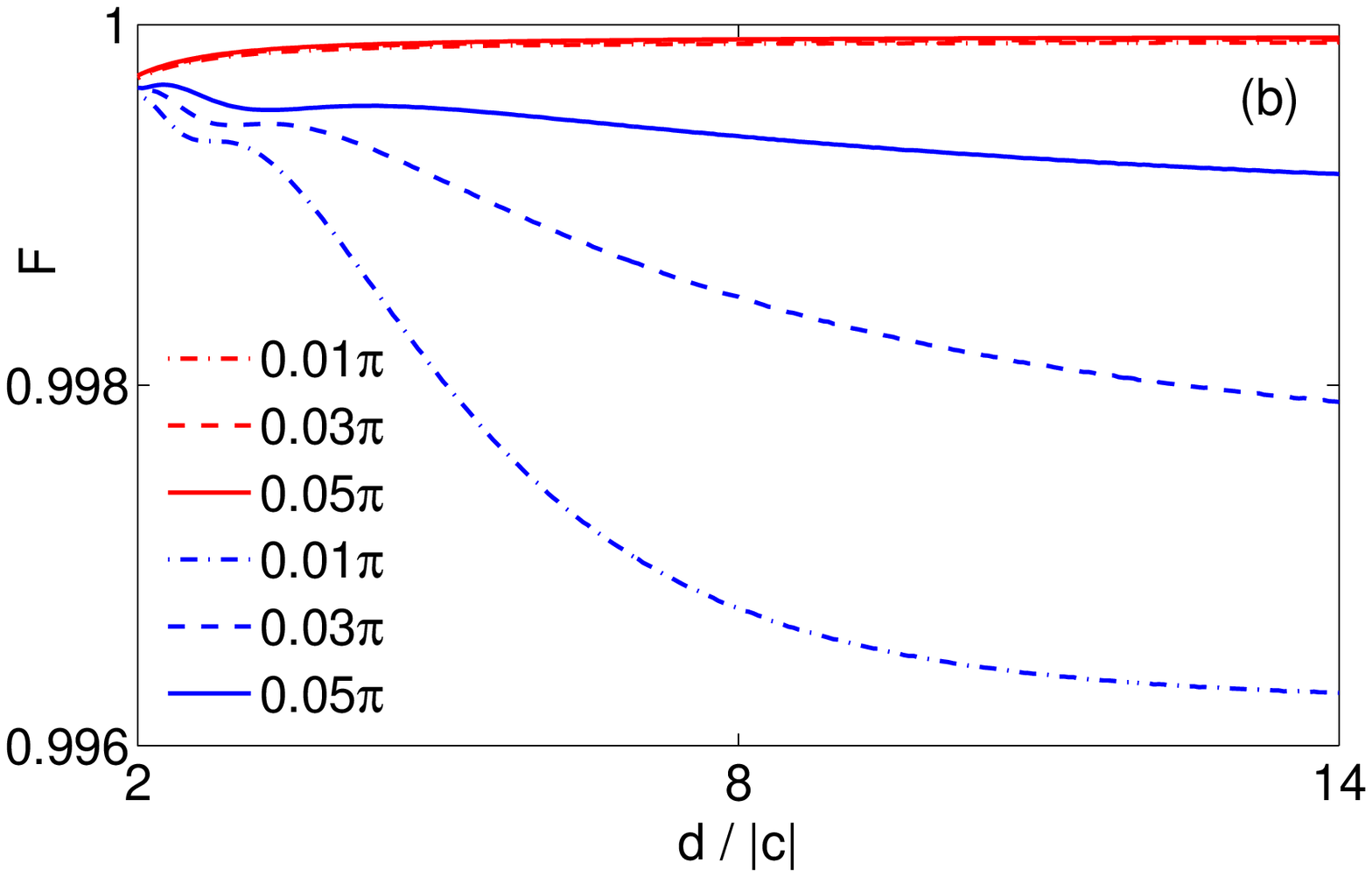}
\caption{(Color online) (a) Fidelity after 1 cycle (blue solid lines), 10 cycles (red dashed lines), 100 cycles (black dash-dotted lines), and 200 cycles (purple dotted lines) under the two-step protocol. For clarity, each curve is shifted up by $0.01$ from bottom to top. The insert shows the zoom-in view near the dip in the main panel. (b) Fidelity after 1 cycle (red lines) and 10 cycles (blue lines) under the four-step protocol for $\tau / \pi = 0.01$ (dash-dotted lines), $0.03$ (dashed lines), and $0.05$ (solid lines). The results show that both the two-step protocol and the four-step protocol are robust, i.e., the fidelity $F$ is close to 1, in the presence of relatively 5\% magnetic pulse errors.
}
\label{fidelity}
\end{figure}

We have assumed the magnetic control pulses are perfect in previous sections, but there are always uncontrollable errors in practical experiments, e.g., the microwave field $\delta$ may have 5\% relative uncertainty~\cite{Zhao14}. Since the timing is pretty accurate in current experiments, we next evaluate the robustness of the two-step and four-step protocols only under the 5\% uncertainty of $\delta$ for many control cycles.

We define the fidelity of a protocol after many control cycles as
\begin{equation}\label{eq:f}
    F = |\langle \vec \xi_i | \vec \xi_f \rangle |^2
\end{equation}
where $|\vec \xi_{i,f}\rangle$ is the initial and the final state of the spin-1 condensate and satisfies $|\langle \vec \xi |\vec \xi \;\rangle |^2 = 1$. The state has three components, $|\vec \xi \;\rangle = (\xi_+, \xi_0, \xi_-)^T$ with $\xi_{\alpha} = \sqrt{\rho_\alpha} e^{-i\theta_\alpha}$ and $\rho_\alpha$ and $\theta_\alpha$ being the fraction and the phase of the component $\alpha$, respectively. The fidelity measures how close the initial and the final states are. The fidelity is 1 for ideal pulses but lower than 1 in the presence of pulse errors. The larger the errors are, the lower the fidelity is. Higher fidelity indicates more robustness of the protocol to pulse errors.

We assume the magnetic field error is distributed with equal probability in the range $[0.95, 1.05] d$ with an average of $d$. For the four-step protocols, the errors for $d_1$ and $d_2$ are independent. We numerically calculate the dependence of the fidelity $F$ on $d$ and show the results in Fig.~\ref{fidelity}(a) for two-step protocols and \ref{fidelity}(b) for four-step protocols.

As shown in Fig.~\ref{fidelity}(a), the fidelity is above 99\% for most $d$, except a special dip near $d/|c|\approx 1.2$ where the period $T$ is most sensitive to the change of $d$ [i.e., the largest derivative of $T$ with respect to $d$ in Fig.~\ref{2step}(b)]. This result manifests that the two-step control protocols are pretty robust under the uncertainty of the magnetic field, if we choose a field away from the dip.

For the four-step protocols, as shown in Fig.~\ref{fidelity}(b), we find that the fidelity is very close to 1, though it decreases as $d$ increasing or $\tau$ decreasing. By taking into account of the requirement of small $\tau$ and large $d$ to better localize the condensate dynamics, a delicate balance between the localization and the robustness is required in practical experiments.

\section{Conclusion}
\label{sec:con}

We propose to to localize the spin mixing dynamics in a spin-1 Bose condensate by periodically applying magnetic pulse sequences, according to the two-step protocol for an initial state with small initial relative phase or the four-step protocol for an arbitrary initial state. Numerical calculations confirm the validity of the proposal for a ferromagnetically interacting spin-1 condensate under the single spatial mode approximation. We further illustrate the robustness of the localization protocol with numerical calculations by assuming 5\% uncertainty of the magnetic pulse amplitude, which might occur in practical experiments~\cite{Zhao14}. Our proposal may be utilized to realize higher precision magnetometers based on spinor BEC~\cite{Vengalattore07, Eto13} or to explore the weak dipolar interaction effects in $^{87}$Rb spin-1 condensates by suppressing the spin dynamics induced by the spin exchange interaction~\cite{Sadler06, Kawaguchi07, Vengalattore08, Eto14, Zhang15}.

\acknowledgments

This work is supported by the National Basic Research Program of China Grant No. 2013CB922003, the National Natural Science Foundation of China Grant No. 11275139, the NSAF Grant No. U1330201, and the Fundamental Research Funds for the Central Universities.


\end{document}